\documentclass[final,3p,times,twocolumn]{elsart/elsarticle}
\usepackage{xspace,relsize,eurosym,color,graphicx,rotating}

\input superbsym

\def\allpixel {all-pixel\xspace}

\journal{Nucl.\ Instrum.\ Methods A}

\begin{document}

\begin{frontmatter}

\title{Design Study for a \superb\ Pixel Vertex Detector}

\author[qmul]{A.~Bevan}
\ead{a.j.bevan@qmul.ac.uk}

\author[ral]{J.~Crooks}
\author[ral]{A.~Lintern}
\author[ral]{A.~Nichols}
\author[ral]{M.~Stanitzki}
\ead{marcel.stanitzki@stfc.ac.uk}
\author[ral]{R.~Turchetta}
\author[ral]{F.~F.~Wilson}
\ead{fergus.wilson@stfc.ac.uk}

\address[qmul]{Queen Mary, University of London, E1 4NS, UK}
\address[ral]{STFC, Rutherford Appleton Laboratory, Chilton, Didcot,
  OX11 0QX, UK}

\begin{abstract}
We present a conceptual design for a low-mass, \allpixel vertex detector using
the CMOS quadruple well INMAPS process, capable of working in the very
high luminosities exceeding $10^{36}\cm^{-2}\sec^{-1}$ that can be
expected at the next generation \epem\ \BFs. We concentrate on the
vertexing requirements necessary for time-dependent measurements that
are also relevant to searches for new
physics beyond the Standard Model. We investigate different
configurations and compare with the baseline designs for the \superb\ and
\babar experiments. 
\end{abstract}


\end{frontmatter}


\section{Introduction}\label{sec:intro}

This document summarises the conceptual design and the supporting physics
studies for an \allpixel vertex detector to be considered at the proposed
next generation high luminosity \epem collider called \superb~\cite{ref:white_det,ref:cdr}. 
We give a short introduction to the physics case for \superb\ and the
motivation for an \allpixel\ vertex detector based on MAPS
technology. We describe the pixel sensor design parameters in
section~\ref{sec:sensors} and the mechanical support structure in
section~\ref{sec:mechanics}. In section~\ref{sec:physics}, we discuss
a number of options for different geometries and use the simulated
\Bz\to\pip\pim decay channel to characterise the performance of the
vertex detector in terms of its time resolution and \B-flavour
tagging efficiency. More information on \superb\ can be found in the
\superb\ Detector `White Paper`~\cite{ref:white_det}, the Conceptual
Design Report~\cite{ref:cdr}, and the proceedings of the SLAC Workshop
on a \sff~\cite{ref:potential}.

\superb\ is a next generation high luminosity \epem\ collider
\BF\ that exploits small beam emittances ($\epsilon_x < 2.5$\nm
and $\epsilon_y < 6$\picom) to deliver high luminosities with moderate
currents of $\sim 2$A and an estimated power consumption of less than 20
MW~\cite{panta_2006,ref:white_acc}. The luminosity will be above
$10^{36}\cm^{-2}\sec^{-1}$, a factor 100 greater than today's
\BFs, allowing a data sample of 75\invab\ to be accumulated
within five years of nominal running.

\superb\ will perform precision tests
of the Standard Model and searches for new physics phenomena. For example
 investigations of the current deviations from the Standard Model predictions
 at the level of $2-3\sigma$ from the existing \BFs, \babar and \belle. 

The ability of \superb\ to look for new physics signals through
indirect searches is complementary to the direct searches
that are underway at the Large Hadron Collider (LHC). \superb\ can
also search for new physics at scales beyond the reach of the LHC.
Precision tests of Charge-Parity-Time (CPT) conservation at \superb
will probe new physics at the Planck scale. CP violation parameters in
\B\ and $D$ decays are sensitive probes of Higgs and Super-Symmetric
(SUSY) particles.  \superb\ can combine information from rare \B\
decays to precisely measure $\tan\beta$ (the ratio of the
Higgs-doublet vacuum expectation values) or the coupling $A$ in
Constrained Minimal Super-Symmetric Models (CMSSM). \superb\ can also
search for charged Higgs particles to a level that exceeds the LHC
search capabilities by a factor of 3-5 over the full range of
$\tan\beta$. Using rare decays \superb\ will be able to measure
flavour couplings in the squark sector to a few percent. \superb\ will
search for Lepton Flavour Violation (LFV) in the decays of the
$\tau$-lepton down to branching fractions of the level of $2\times
10^{-10}$; SUSY Grand Unified Theory (GUT) models, using constraints
from the current \Bs\ mixing and phase measurements from the Tevatron,
predict that LFV decays could exist with branching fractions of a few
$10^{-8}$~\cite{ref:white_physics}.

The \superb\ detector is a refinement on the \babar design and described in
detail in~\cite{babarnim}. As \superb\ will operate at a lower centre of
mass boost, $\beta\gamma=0.2375$ (6.7\gev$e^{-}$ beam against a
4.2\gev\ep\ beam), the average spatial vertex separation between the two
decaying \B\ mesons is $\langle\Delta z\rangle \approx \beta\gamma c
\tau_B = 110\mum$, a factor of two smaller than
\babar.

For its vertex detector, \superb\ plans to use a similar design as the
\babar detector, with five layers of silicon strips with a radius
between 3 and 15\cm, but with an additional inner layer of striplets at a
radius of $\sim 1.6$\cm to the beam (Layer 0). The Layer 0 is the
biggest challenge for the design of the \superb\ silicon vertex
detector. \superb\ simulations suggest a maximum hit rate in Layer 0 of
$\sim 100 \MHz/\cm^{2}$. This high data rate and the associated power
consumption require an active cooling solution while still maintaining
a small material budget. The radiation damage at \superb\ is also an
issue with the expected radiation dose for Layer 0 of $\sim 10$ Mrad,
corresponding to 10$^{13}$ neutrons. Up to now, there have been
several proposals to realise Layer 0, based on either LHC-style hybrid
pixels~\cite{hybrids} or striplets~\cite{striplets}.

Our proposal is to replace the current six layer design of \superb
with a six layer solution using pixels for all the layers with a pixel
size of $50 \times 50$ \mum, corresponding to 2500 hits per second per
pixel for Layer 0. This is made possible by using Monolithic Active Pixel Sensor
(MAPS) technology \cite{MAPSFORPP}. To simplify assembly and testing
we are planning to use a long barrel design with the sensors mounted on a
stave structure with integrated cooling. This will allow easy exchange
of a stave in case of problems e.g. a massive beam incident. An
homogeneous detector using the same solution everywhere will make
construction and maintenance easier. However an \allpixel solution for
the vertex detector poses a challenge as both high granularity and a
low material budget are important.

\section{Sensor Design}\label{sec:sensors}

This \superb\ pixel design study is based on circuitry already
developed for the Tera-Pixel Active Calorimeter (TPAC)
chip~\cite{Ballin:2008db}. This design is intended to be used in a
digital electromagnetic calorimeter for the International Linear
Collider (ILC)~\cite{Watson:2008zzd,Crooks:2008zz}. Also considered is
the more advanced four-transistor (4T) architecture already explored
in the Fortis chip~\cite{4ttwepp09}. Each \superb\ chip is assumed to
be $2.5\times 2.5$ \cm
in size, and four chips are stitched together to make a module with 
an active area of $2.5 \times 10$ \cm, and a pixel size of 
$50 \times 50$ \mum. The hits in the silicon
pixels are also to be used in \dedx\ measurements, so an ADC with 4-5
bit resolution is required.  The readout is based on a column
architecture with the electronics at the short end of the module. Each
chip has one million pixels and each module therefore has four million pixels with 2000
columns and 500 pixels per column.  The raw data rate per module is
10 \gbysps, assuming a hit rate of 2500 hits per second per pixel and
a data size of 32 bits per pixel (row/column address plus time stamp
and 5-bit ADC value). This illustrates the need for on-chip data
reduction. As timing information is required, the chips will be
sampled every 500\ns, providing enough granularity in time to provide
efficient pattern recognition in Layer 0.

The TPAC chip has been designed using the 180\nm\ CMOS quadruple well
INMAPS process~\cite{Ballin:2008db}, which includes a deep p-well
implant. This allows the use of full CMOS capabilities, as the
n-wells of the PMOS devices are now shielded by the deep p-well. This
is a significant advantage to previous MAPS devices which were limited
to the use of NMOS technology because of the parasitic charge
collections of the PMOS n-wells.

\begin{figure}[!ht]
\begin{center}
 \includegraphics[width=0.95\columnwidth]{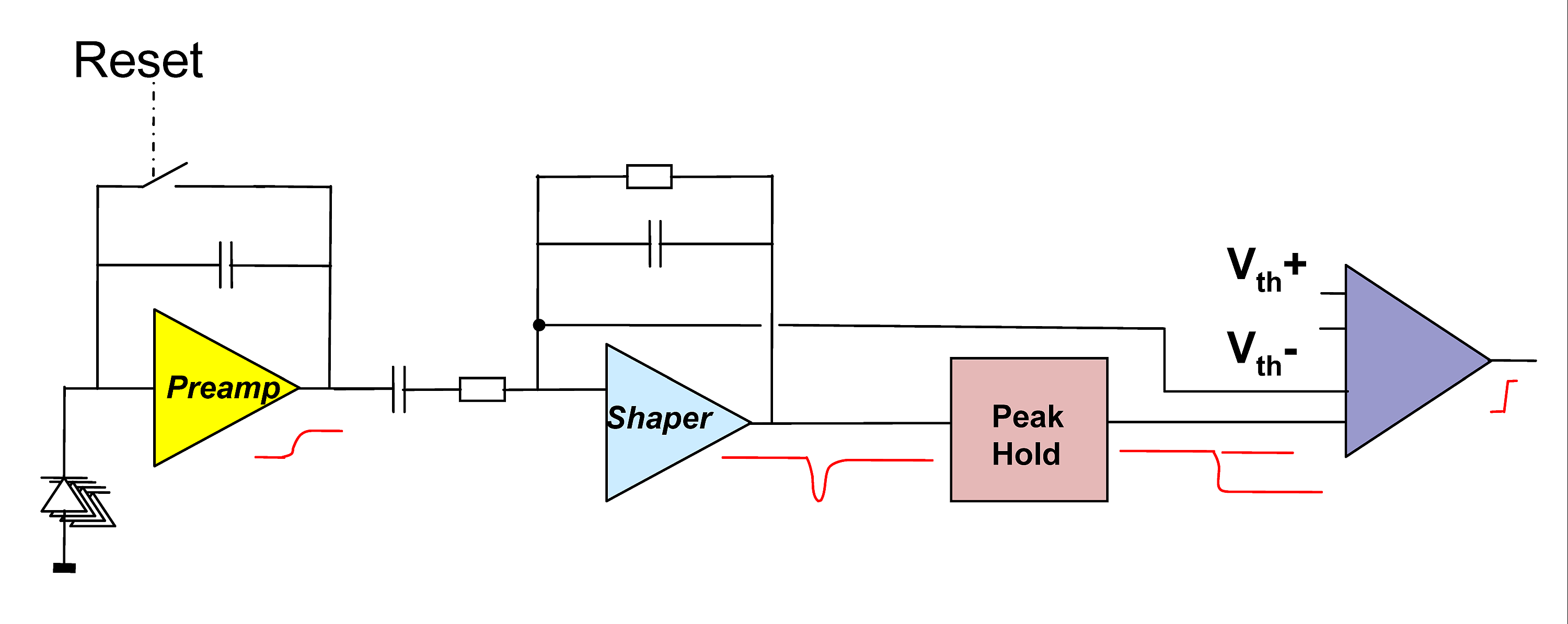}
\caption{\label{fig:tpac_for_superb}Schematic of the TPAC pixel for \superb,
showing the preamplifier, the shaper and the comparator and peak-hold blocks.}
\end{center}
\end{figure}

The original TPAC pixel contains a preamplifier, a shaper and a
comparator~\cite{Ballin:2008db}. The pixel only stores hit information
in a 14-bit ``Hit Flag". The pixel itself runs without a clock and the
timing information is provided by the logic querying the ``Hit
Flag". For the \superb\ application, a peak-hold latch was added as shown in
Fig.~\ref{fig:tpac_for_superb} to store the analog information as well. 
The chip is organised in columns with
a common ADC at the end of each column. The ADC is realised as a 5-bit
Wilkinson ADC using a 4~\MHz\ clock. For the predicted
time-stamping resolution of 500\ns, there are on average 1.25 pixels
hit in a column. The simulated power consumption for each individual
pixel is less than $12\muWatt$ and the power consumption per module is
smaller than 12 W.

\begin{figure}[!ht]
\begin{center}
\includegraphics[width=0.95\columnwidth]{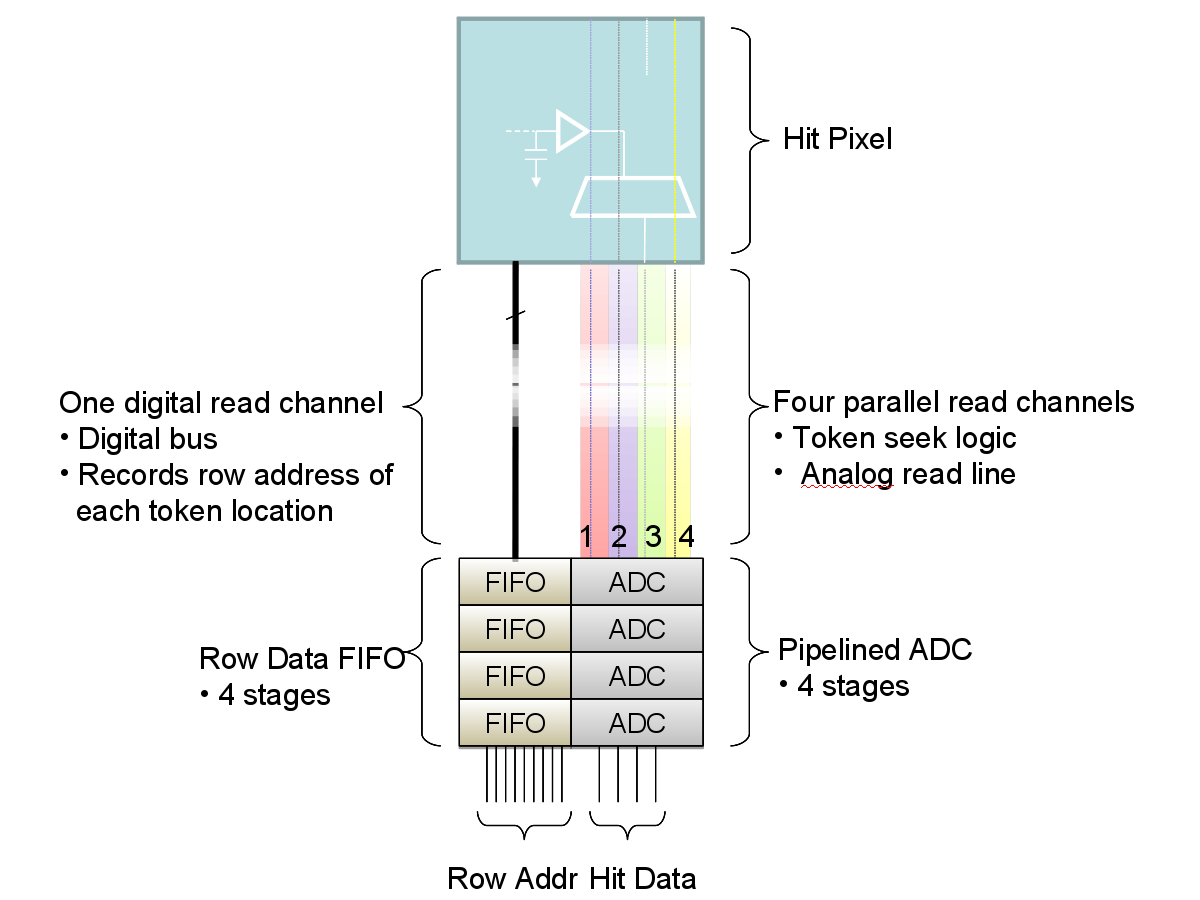}
\caption{\label{fig:readout_scheme1} The basic architecture of one column in the
TPAC for the \superb\ chip, showing the digital and analog readout lines and the pipelined ADC
and FIFO at the end of the column.}
\end{center}
\end{figure}

The column logic constantly queries the pixels for their ``Hit Flag"
(Token Seek logic) but only digitises the information for the pixels
with a ``Hit Flag" high. This saves both space and reduces the power
usage. Since the speed of the chip is limited by the ADC, the Token
Seek logic also increases the readout speed. Both the address of the
pixel being hit and its ADC output are stored in a FIFO at the end of
the column. This is shown in Fig~\ref{fig:readout_scheme1}.

\begin{figure}[!ht]
\begin{center}
\includegraphics[width=0.95\columnwidth]{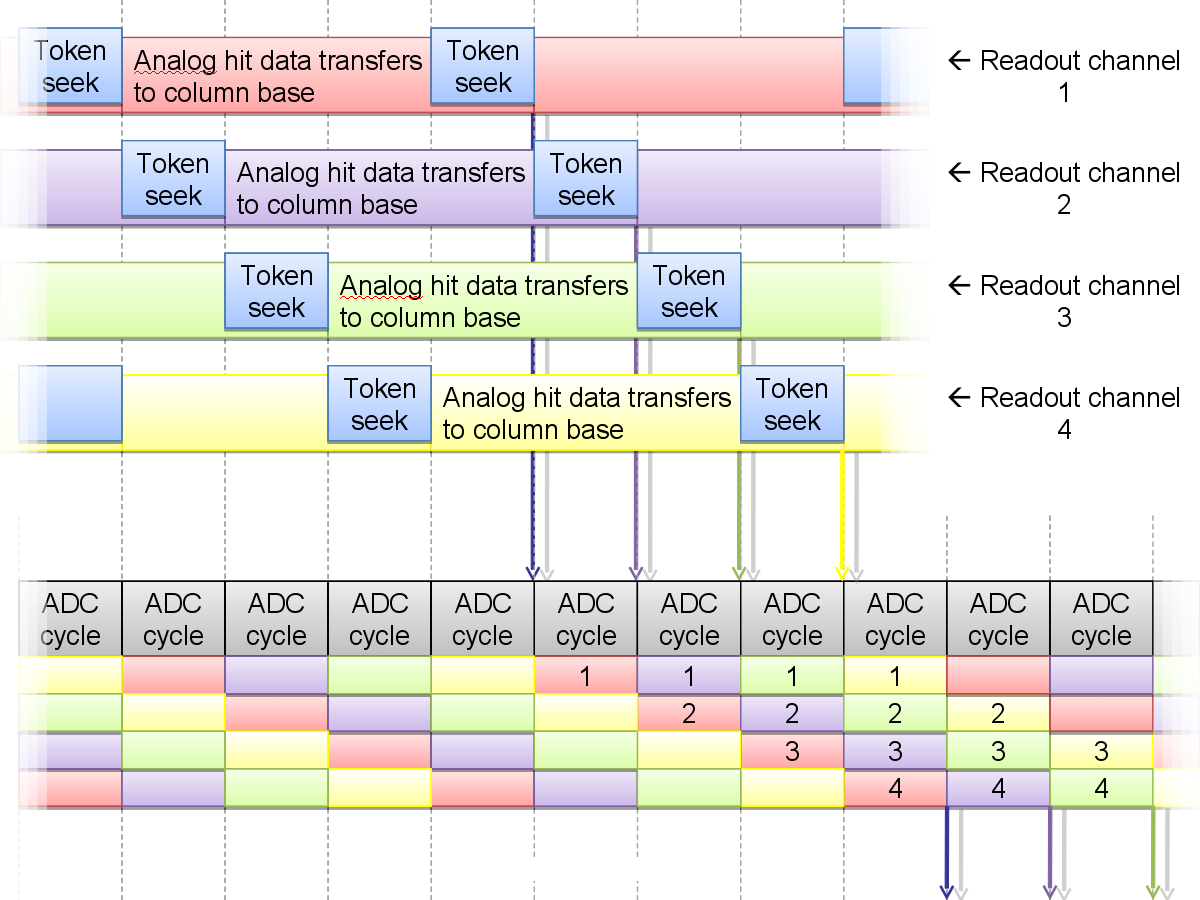}
\caption{\label{fig:readout_scheme2} The timing diagram for the four-stage pipelined ADC.}
\end{center}
\end{figure}

As the analog charge transfer from the individual pixels to the end of
the column is slow, the ADC uses a pipelined architecture with 4 analog input lines to
further increase throughput of the ADC. The pipeline design is illustrated in
Fig.~\ref{fig:readout_scheme2}.

\begin{figure}[!ht]
\begin{center}
\includegraphics[width=0.95\columnwidth]{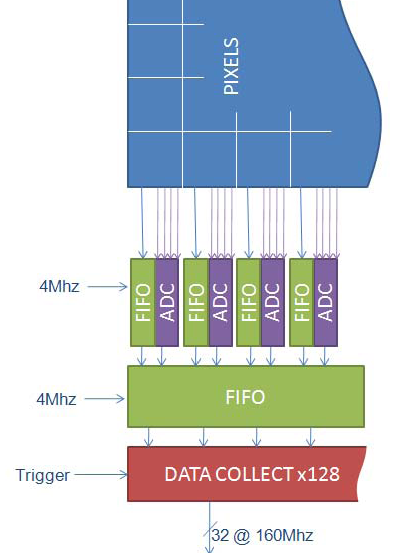}
\includegraphics[width=0.95\columnwidth]{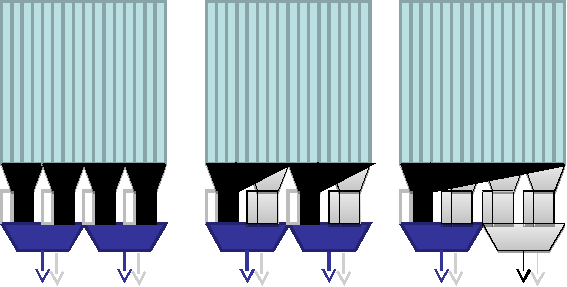}
\caption{\label{fig:readout_scheme3} The overall column design
including the Event FIFO, the data sparsification logic using the
Level 1 Trigger information and the high speed link to the external
DAQ and the readout of the layers (bottom). The readout can be
multiplexed in the outer layers to save power.}
\end{center}
\end{figure}

One of the main bottlenecks is reading the data from the module. The minimum
irreducible event rate in the detector is expected to be 50~\kHz\ from Bhabha
scattering and 20~\kHz\ from physics production~\cite{ref:white_det}. A
hardware-based Level 1 trigger will receive data at 4~\MHz, using information
from the drift chamber and calorimeter, and will accept events for further
processing at a design limit of 150~\kHz. Readout of the vertex detector will
only happen on receipt of the Level-1 Accept signal and this will reduce the
data rate by a factor of 10. This also minimises power and services
required, as shown in Fig.~\ref{fig:readout_scheme3}.

For the outer layers, the hit rate is much smaller and the timing
requirements can be relaxed. To save power, the ADCs in these layers
will be multiplexed to handle more than one column in the sensor. A
multiplexing scheme for this device is also illustrated in
Fig.~\ref{fig:readout_scheme3}.

\begin{figure}[!ht]
\begin{center}
\includegraphics[width=0.95\columnwidth]{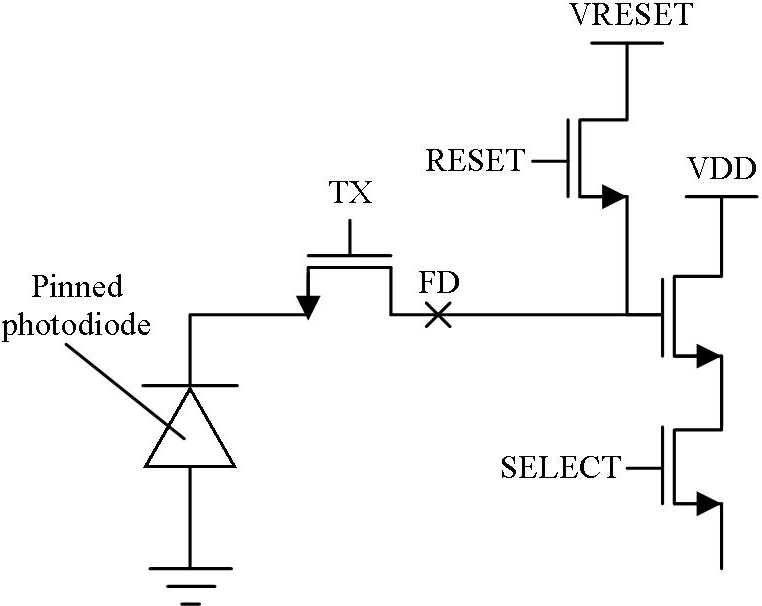}
\caption{\label{fig:4Tpixel} The basic four transistor 4T pixel with the pinned Diode, 
the Transfer gate and the floating diffusion node.}
\end{center}
\end{figure}

An alternative to the TPAC chip is a four transistor (4T) structure as
has already be demonstrated by the Fortis chip~\cite{4ttwepp09}. This
would allow ultra-low noise performance on a level of less than eight
electrons, a much smaller power consumption of about
2$\mu$W per pixel and the possibility of pixel sizes less than $50
\times 50$ \mum. The limiting factor for this approach is the speed of
the charge transfer in the 4T structure (see Fig.~\ref{fig:4Tpixel}),
which is currently limited to 1-2 $\mu$s. If Correlated Double
Sampling (CDS) is used to reduce the noise, a second readout is
required and so the total readout time is 2-4\mus which is close to
the trigger accept limit. Fast charge transfer structures would need
to be added in order to satisfy the \superb\ timing requirements.

The TPAC chip uses standard industry processes and this significantly
 reduces the production costs. Yields of at least 60~\% are achievable for these
 sensors. Even when set-up costs are taken into
 account, it is still financially feasible to regularly replace the
 sensors in the inner layer, eliminating the need for extreme
 radiation hardness.


\section{Mechanical Design}\label{sec:mechanics}

\begin{figure}[!ht]
\begin{center}
 \includegraphics[width=0.95\columnwidth]{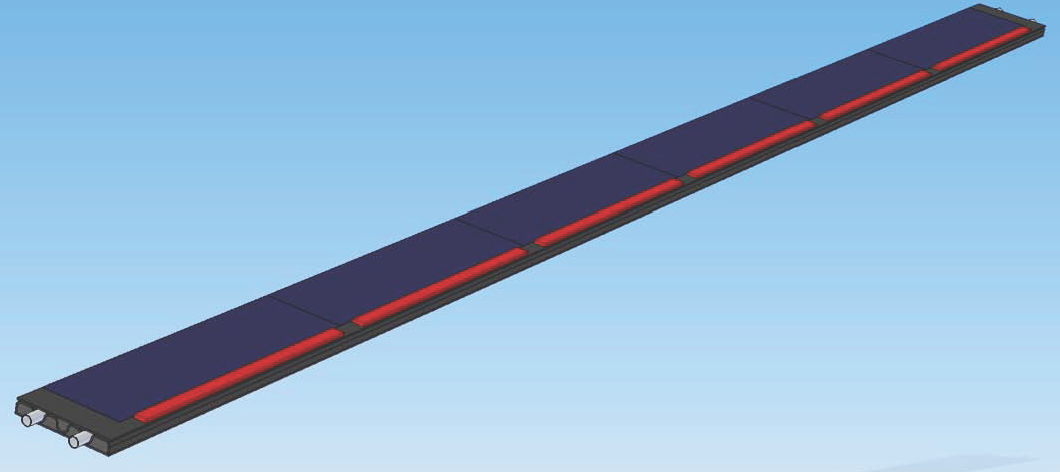}
 \includegraphics[width=0.95\columnwidth]{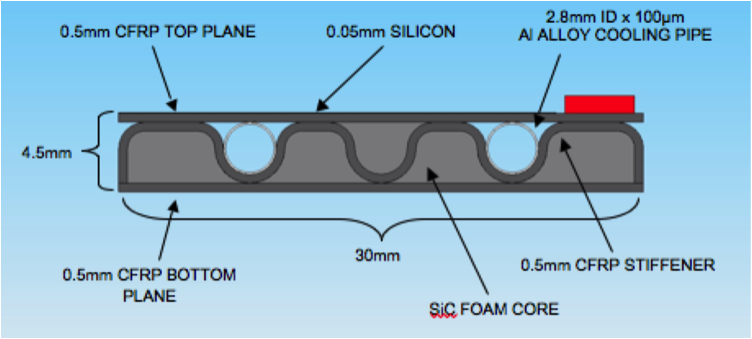}
\end{center}
\caption{The \superb\ stave design (top), with a
  cross-sectional view of the stave (bottom) showing the Carbon Fibre
  stiffener (CFRP), Silicon Carbide (SiC) foam and channels for water
  cooling. The dark blue area corresponds
to the sensor, and the red strips are the on-stave electronics. The
  stave must be rigid over a distance of 60\cm.}
\label{fig:stave}
\end{figure}

The mechanical concept outlined here is for a barrel detector geometry
built from units of modules arranged on staves. Each sensor module consists of 4 chips and
has a size of $2.5 \times 10$ \cm with a thickness of 50 \mum. This forms the
irreducible unit that the remainder of the detector is constructed
from. Figure~\ref{fig:stave} shows the stave concept for the
outermost layer. The cooling, power and readout is connected to the
stave at one or both ends, to facilitate installation and replacement
should modules fail for any reason. The coolant is assumed to be
water, and in order to maintain suitable operational conditions we
need to be able to extract about $10\Watt$ of power from each sensor module,
which corresponds to $0.4\Watt/\cma$. The cooling pipes are assumed
to be made from thin walled Al tube, however it may be possible to try
alternative pipe materials if there is a significant reduction in the
material budget.

The material budget for this design is shown in
Table~\ref{tbl:stavematerial}. The sensor consists of INMAPS active
material with a 50\mum thick layer of implanted silicon; a carbon
fibre support structure (CFRP); an Aluminium cooling pipe; coolant
(water); and a SiC Foam (4\% density) modeled as an equivalent amount
of Carbon Fibre. The majority of the material in the stave is Carbon
Fibre. This design is rigid, with a 250 \mum sag over a distance of
60\cm, which is the maximum distance between support structures in the
outer layer. The total material budget conservatively corresponds to a
total radiation length of $X/X_0= 1.147\%$. This is adequate in the
outer layers but has a detrimental effect in the inner layers. The
inner most layer(s) could benefit from a significantly lower mass
support structure. An ultra-low mass support for Layer 0 using a
peek-carbon fibre support layer with $X/X_0=0.17\%$ has been
proposed~\cite{Bosi} and we are considering this design as an alternative.

\begin{table}
\caption{The material content of a stave (cross-sectional view) giving
  the equivalent thickness averaged over the stave and the corresponding 
radiation length.
}\label{tbl:stavematerial}
\begin{center}
\begin{tabular}{llrc}
\hline 
\hline
Substance & Thickness & Radiation   & Radiation \\ 
          &  (mm)      & length (mm) & length (\%)  \\ \hline
Silicon      & 0.05 &   94 & 0.053 \\
CFRP         & 1.752 &  240 & 0.730 \\
Al           & 0.0623 &  89 & 0.069 \\
Water        & 0.41     &  360 & 0.114 \\
SiC          & 1.81 & 1000 & 0.181 \\
Total        & &      & 1.147 \\ 
\hline 
\hline
\end{tabular}
\end{center}
\end{table}

\begin{figure}[!ht]
\begin{center}
 \includegraphics[width=0.95\columnwidth]{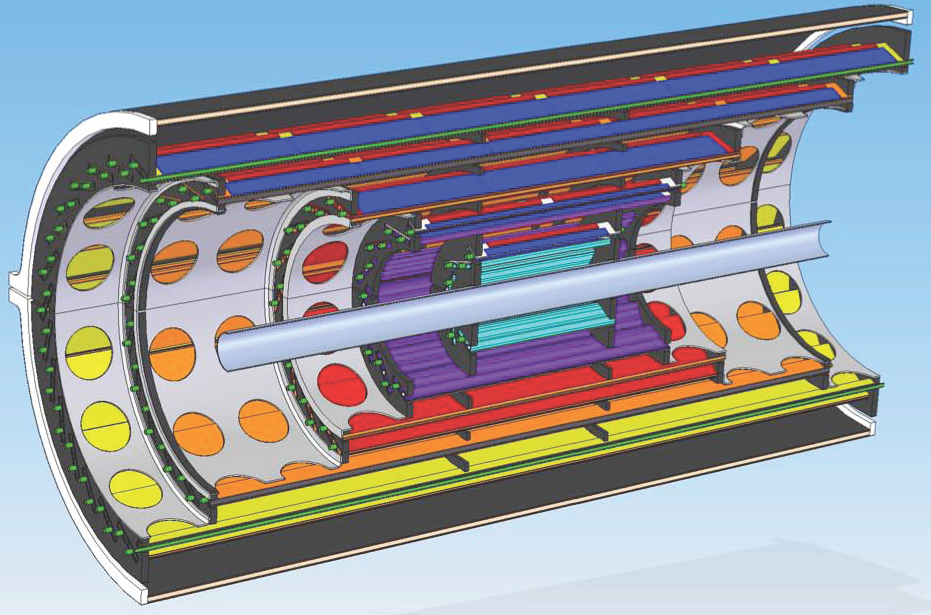}
 \includegraphics[width=0.95\columnwidth]{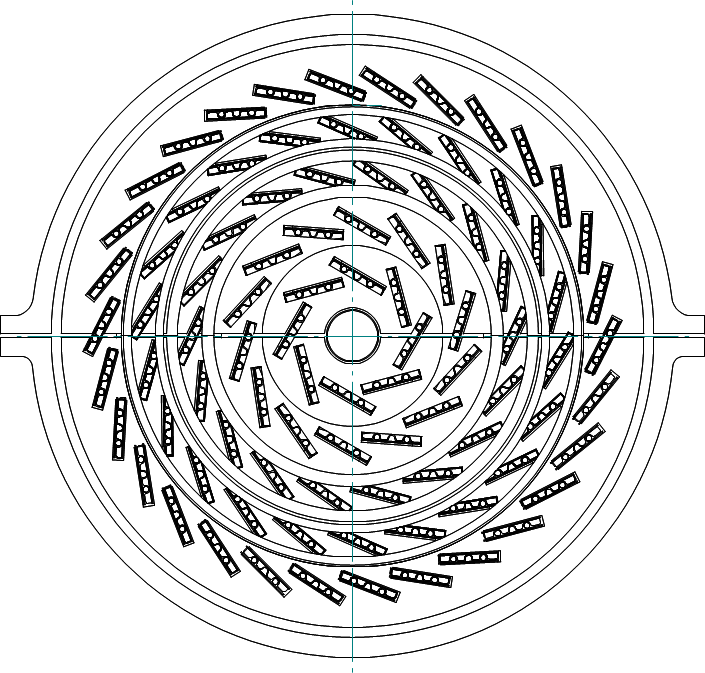}
\end{center}
\caption{The \superb\ vertex detector design: a half-clamshell of
the detector (top), and a beams-eye-view showing the pin-wheel
structure of the detector (bottom).}
\label{fig:svt}
\end{figure}

The baseline geometry for the \superb\ vertex detector assumes a lamp-shade
design similar like the Babar vertex detector design \cite{babarnim},  where the
ends of the outer layers are angled with respect to the beam line to reduce the
multiple scattering for low angle tracks. Here we have considered a long barrel
detector with many staves joined together in two detector halves. These two
halves are brought together in the final detector.  This has the advantage of
uniform modules.  However as low angle tracks will pass through more material in
a long barrel detector, this design can suffer from higher multiple scattering
for  such tracks. However the depth of the charge collection in the sensor is
sufficiently small that this is not the major error on the track measurements.
Fig.~\ref{fig:svt} shows a cut-away view of one half of the detector for a five
layer design. Also shown is a sectional view of the detector illustrating the
overlapping layout of the individual staves.

\section{Physics Studies}\label{sec:physics}

We have studied the performance of the \superb\ detector with different
geometries of a vertex detector, ranging from the \superb\ baseline,
through to a four or six layer \allpixel detector with a realistic
material budget for the support structure for all layers. The decay
channel chosen for these studies is \upsbzbz, where one of the \Bz
decays through the channel \Bz\to\pip\pim, as this tests the ability
to reconstruct tracks and identify the positions of the two \B\ mesons
when they decay. A critical aspect of the detector is the ability to
compute the proper time difference \deltat\ between the decays of the
two \B\ mesons in the event, as computed from the reconstructed
spatial separation \deltaz between the vertices and the known boost
$\beta\gamma$ of the experiment. The signal reconstruction efficiency
for these events and the resolution on \deltat\ are used to compare a
number of different vertex detector geometries, with the rest of the
detector assumed to be the baseline as outlined by the \superb\
CDR~\cite{ref:cdr}.  A good \deltat\ resolution is vital for \superb\
to measure accurately the \CP\ asymmetry parameters $S$ and $C$, which
in turn can be used to search for new physics for a number of rare
final states of a \B\ meson. The vertex detector must also be able to
identify the flavour of the other \B\ in the event (the ``tagged \B'')
in order to make time-dependent \CP\ measurements, as well as a number
of other new physics searches that will be performed at \superb. This
requires in particular efficient reconstruction of pions with low
transverse momentum ($\sim 100\mevc$). This benchmark decay is
therefore a validation of both searches for new physics and Standard
Model calibration measurements, and is a good indicator of performance
across a wide-range of physics channels.


All studies documented here use the \superb\ Fast
Simulation V0.1.1~\cite{ref:white_det}. Events are selected using the same
analysis criteria as the \babar\ \Bz\to\pip\pim
analysis~\cite{babarpipianalysis}. The simulation assumes a boost
$\beta\gamma=0.28$, which is slightly larger than the boost currently
foreseen.

We have considered the following geometries:

\begin{enumerate}
 \item The baseline \superb\ geometry [\superb\ Baseline].
 \item The \babar geometry with the PEP-II beam conditions [\babar Baseline].
 \item A 6-layer all Hybrid Pixel detector (layers: $0 - 5$) [Hybrid Pixels].
 \item A 4-layer all Hybrid Pixel detector (layers: $0,1,4,5$) [Hybrid Pixels-4A].
 \item The baseline \superb\ geometry with an INMAPS detector for Layer 0 [INMAPS-L0].
 \item A 6-layer \allpixel INMAPS detector (layers: $0 - 5$) [INMAPS].
 \item A 4-layer \allpixel INMAPS detector (layers: $0,1,4,5$) [INMAPS-4A].
 \item A 4-layer \allpixel INMAPS detector with layers at 1.6\cm, 5\cm,
 10.2\cm and 14.2\cm radii) [INMAPS-4B].
 \item A 6-layer \allpixel INMAPS detector with a low mass Layer 0 support (layers: $0 - 5$) [INMAPS-LL0].
\end{enumerate}

The first two geometries serve as baselines in order to evaluate the
performance expected from an \allpixel INMAPS based silicon detector.
The two geometries using Hybrid Pixel technology require more material
than INMAPS and are used only to provide a bound for any
possible effects of underestimation of the material budget for
the cooling and the support structure. Geometries
five to eight cover various alternative layouts for the number and
radii of an \allpixel\ INMAPS detector with each layer having the
material budget shown in Table~\ref{tbl:stavematerial}. The last
geometry uses the INMAPS material budget for the outer layers and
the ultra-low mass support for Layer 0 ($X/X_0=0.17\%$).

\begin{figure}[!ht]
\begin{center}
  \includegraphics[width=0.95\columnwidth]{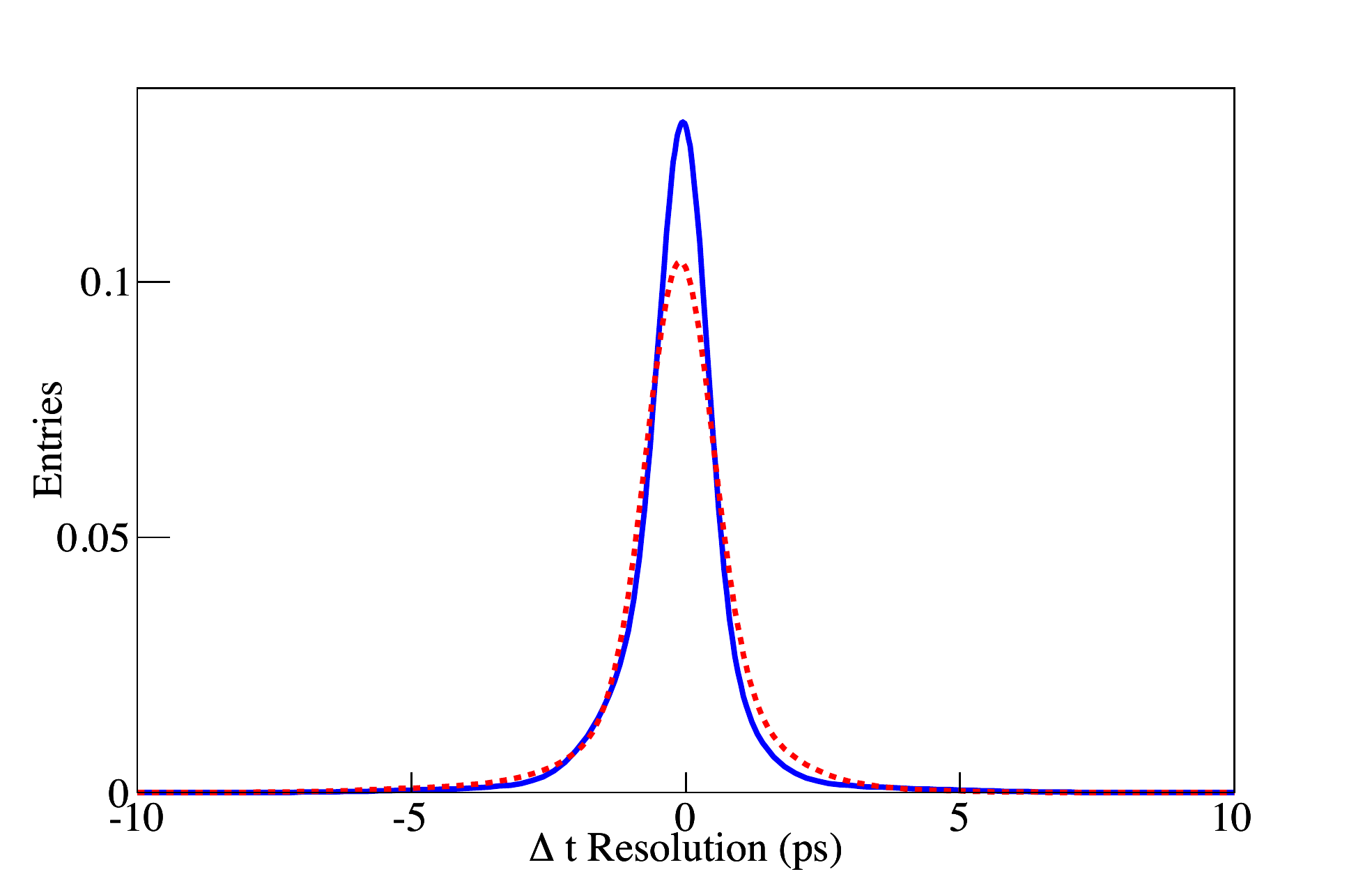}
\end{center}
\caption{Comparison between the \deltat distribution for \superb\
  baseline configuration (geometry 1, dashed) and the six layer,
  low-mass, \allpixel INMAPS configuration (geometry 10, solid). Both
  distributions are fitted with a triple Gaussian function.}
\label{fig:resocomparison}
\end{figure}

Table~\ref{tbl:pipi:reso} summarises the \deltat resolution for the
\Bz\to\pip\pim\ benchmark channel for the nine geometries considered. 
The \deltat\ distribution for events with both \Bz\
decays reconstructed is fitted with a function consisting of three
Gaussians. The table shows the Root Mean Squared (RMS) and the
Full-Width Half-Maximum (FWHM) of the \deltat\
distribution. $f_{\mathrm{core}}$ represents the fraction of the
fitted function that is assigned to the Gaussian with the smallest
width $\sigma_{\mathrm{core}}$. Smaller widths and larger
$f_{\mathrm{core}}$ indicate better performance. The \allpixel
detector geometries are clearly compatible with the \superb\ baseline
in terms of their resolution (geometries five to eight) and such a
detector with a low mass support structure for Layer 0 (geometry 9)
is superior to this, and gives a similar resolution to
as the existing \babar\ detector (geometry 2). A comparison of the
resolution function obtained for this sample, with respect to the
\superb\ baseline is shown in
Figure~\ref{fig:resocomparison}. Overall, the \allpixel detector in
its various configurations is between 12\% and 20\% better than the
baseline in terms of the RMS or FWHM, while still maintaining the same
signal reconstruction efficiencies and energy resolutions.
 
\begin{table*}
\caption{Comparison of \deltat resolution for the decay
 mode \Bz\to\pip\pim in terms of Root Mean Squared (RMS) and
  Full-Width Half-Maximum (FWHM) for different geometries. For each geometry, the
  \deltat distribution is fitted with a triple Gaussian
  function. $\sigma_{\mathrm{core}}$ and $f_{\mathrm{core}}$ are the
  width and fractional area of the core Gaussian, respectively.
}\label{tbl:pipi:reso}
\begin{center}
\begin{tabular}{llccccc}
\hline 
\hline
 & Configuration              & RMS (\ps) & FWHM (\ps)   &
$\sigma_{\mathrm{core}}$ (\ps) & $f_{\mathrm{core}}$ \\ \hline
1 & \superb\ Baseline & $1.232 \pm 0.007$ & 1.44 & $0.692\pm 0.008$ & $0.801\pm 0.008$\\
2 & \babar Baseline  & $1.087 \pm 0.010$ & 1.33 & $0.561\pm 0.015$ & $0.721\pm 0.030$\\
3 & Hybrid Pixels    & $1.259 \pm 0.001$ & 1.54 & $0.635\pm 0.024$ & $0.634\pm 0.043$\\
4 & Hybrid Pixels-4A & $1.249 \pm 0.011$ & 1.49 & $0.537\pm 0.022$ & $0.550\pm 0.037$\\
5 & INMAPS-L0        & $1.163 \pm 0.010$ & 1.40 & $0.551\pm 0.002$ & $0.627\pm 0.039$\\
6 & INMAPS           & $1.227 \pm 0.011$ & 1.42 & $0.519\pm 0.036$ & $0.627\pm 0.066$\\
7 & INMAPS-4A        & $1.212 \pm 0.011$ & 1.32 & $0.505\pm 0.050$ & $0.636\pm 0.090$\\
8 & INMAPS-4B        & $1.209 \pm 0.011$ & 1.29 & $0.501\pm 0.024$ & $0.626\pm 0.042$\\ 
9 & INMAPS-LL0  & $1.089 \pm 0.010$ & 1.14 & $0.427\pm 0.027$ & $0.598\pm 0.056$\\
\hline 
\hline
\end{tabular}
\end{center}
\end{table*}

We have computed the efficiency for identifying the flavour of the \B\
decays (\Bz\ or \Bzb) as a function of six pre-defined tagging
categories that correspond to different types of events. Each
category uses information from the reconstructed tag side \B\ to
identify the flavour of the other \B. In order of increasing mis-tag
probability they are: the ``Lepton'' category using fast leptons
($e,\mu$); ``Kaon 1'' using tracks clearly identified as charged kaons
through particle identification (PID) techniques; ``Kaon 2'' using
tracks with a lower probability of being a kaon; ``Pion'' using tracks
clearly identified as charged pions; the ``Other'' category using other
information of a lower quality; and ``Untagged'' which represents all events
which do not fall into the previous five.

The best single metric for comparing the geometries is the total
tagging efficiency $Q$. However, this requires information on the
dilution factor for each tagging category which is not available for
the \superb\ simulations. As an alternative we look at the efficiencies
for the individual categories. The Lepton flavour tag is the most
important category as it has the lowest mis-tag fraction but the other
categories are also vital contributors to the overall efficiency.

Table~\ref{tbl:pipi:tageff} summarises the tagging efficiencies for
the different detector configurations. The efficiencies for
geometries two to nine are given relative to the \superb\ baseline
design. This accounts for differences between the tagging algorithm
implemented in this simulation and numbers reported elsewhere by
\babar and \superb. From this table we can see
that the \superb\ baseline lepton tag efficiency is consistent with the
\babar one but the Pion and Other tagged categories are $\sim 16\%$
lower in efficiency and the \superb\ baseline will have $\sim 15\%$
more untagged events. The four layer detector geometries reduce the
efficiency for lepton tagged events by $\sim 30\%$ from the
baseline. The INMAPS detector has an efficiency comparable with the
\superb\ baseline for lepton tagged events. There is some small
variation in the untagged category efficiency as a function of
geometry, however this is typically at the 0.4\% level.

\begin{table*}
\caption{The flavour tag fractions for the different tagging
  categories based on a preliminary tagging algorithm computed for
  each geometry. The fractions for geometries two to nine are presented relative to the \superb\
  Baseline.}
\label{tbl:pipi:tageff}
\begin{center}
{\small
\begin{tabular}{llcccccc}
\hline\hline
 & Configuration & Lepton & Kaon 1 & Kaon 2 & Pion & Other & Untagged\\ \hline
1 & \superb\ Baseline  & 0.039$\pm$0.002 & 0.108$\pm$0.002 & 0.128$\pm$0.003 & 0.176$\pm$0.003 & 0.110$\pm$0.002 & 0.440$\pm$0.004\\
\hline
2 & \babar Baseline   & 0.949$\pm$0.054 & 1.028$\pm$0.036 & 1.055$\pm$0.030 & 1.142$\pm$0.025 & 1.164$\pm$0.031 & 0.884$\pm$0.015\\
3 & Hybrid Pixels     & 0.923$\pm$0.056 & 0.981$\pm$0.038 & 1.047$\pm$0.030 & 1.006$\pm$0.028 & 0.982$\pm$0.037 & 1.000$\pm$0.014\\
4 & Hybrid Pixels-4A  & 0.718$\pm$0.071 & 1.000$\pm$0.037 & 0.922$\pm$0.034 & 0.915$\pm$0.031 & 0.918$\pm$0.040 & 1.103$\pm$0.012\\
5 & INMAPS-L0         & 0.897$\pm$0.057 & 1.009$\pm$0.037 & 1.023$\pm$0.031 & 0.983$\pm$0.029 & 0.936$\pm$0.039 & 1.023$\pm$0.013\\
6 & INMAPS            & 0.923$\pm$0.056 & 1.009$\pm$0.037 & 1.016$\pm$0.031 & 0.989$\pm$0.029 & 0.982$\pm$0.037 & 1.009$\pm$0.014\\
7 & INMAPS-4A         & 0.692$\pm$0.074 & 1.000$\pm$0.037 & 0.945$\pm$0.033 & 0.943$\pm$0.030 & 0.945$\pm$0.038 & 1.080$\pm$0.013\\
8 & INMAPS-4B         & 0.769$\pm$0.067 & 1.000$\pm$0.019 & 1.000$\pm$0.023 & 1.000$\pm$0.017 & 1.000$\pm$0.018 & 1.021$\pm$0.009\\
9 & INMAPS-LL0        & 0.872$\pm$0.059 & 1.028$\pm$0.036 & 0.969$\pm$0.032 & 1.006$\pm$0.028 & 1.000$\pm$0.036 & 1.011$\pm$0.014\\
\hline
\hline
\end{tabular}
}
\end{center}
\end{table*}

\section{Conclusion}\label{sec:conclusion}

These initial studies suggest that a four layer \allpixel INMAPS
vertex detector will be as performant as the current \superb\ baseline
geometry. A low mass support for Layer 0 would further offset any
performance degradation obtained by halving the boost of the centre of
mass system required by the power budget of the \superb\ accelerator. A
six layer \allpixel INMAPS vertex detector using this support
structure for the inner layer would have a \deltat resolution
comparable to \babar.  The energy resolution and event reconstruction
efficiency for \Bz\to\pip\pim is independent of the amount of material
in the detector but it will be important to study the effect of
material on final states with neutral particles as a complementary
input to the detector optimisation. In addition more studies are
required for time-dependent CP analyses, charm mixing, and other
channels that depend highly on slow pion efficiency.

The long barrel design can be improved by combining the support
structure for the outer layers four and five and so reducing the
amount of material required. A lamp-shade structure, to reduce
multiple-scattering at low angles, can also be designed without the
need to resort to non-standard industry chip dimensions. Both concepts
are under study.

This R\&D effort has produced a viable pixel
vertex detector design for the \superb\ detector that meets the
challenging demands placed on the system in terms of rate and delivers
superior physics performance when compared to other viable alternative
technologies.  This design concept has been developed with uniformity
in mind.  Further R\&D is required to realise a prototype
module and stave structure for further tests. Preliminary tests with
X-rays have produced satisfactory results for the existing TPAC layout
but the design would benefit from further radiation tests to confirm
that the sensors will perform as expected in terms of bulk damage when
exposed to $\sim 10$ Mrad radiation dose. However, the low production
costs of the sensor offer the possibility of a design with a
disposable inner-layer that is replaced each year.

\section{Acknowledgements}

This work has been supported by the Science and Technology Facilities 
Council (STFC).

\bibliographystyle{model1-num-names}

\end{document}